\documentclass[12pt]{article}
\usepackage{setspace}
\usepackage{amsmath}
\usepackage{graphicx}
\usepackage{cases}
\usepackage{subfig}

\usepackage{natbib}
\usepackage{psfrag}
\usepackage{amssymb}
\usepackage{color,graphics,epsfig}
\usepackage{appendix}
\usepackage{fancyhdr}
 
\def\be{\begin{equation}}
\def\ee{\end{equation}}
\def\bea{\begin{eqnarray}}
\def\eea{\end{eqnarray}}
\def\bml{\begin{mathletters}}
\def\eml{\end{mathletters}}

\newcommand{\tc}{\textcolor}

\begin{document}

\title{Response of polygenic traits under stabilising selection and mutation when loci have unequal effects}

\author{Kavita Jain$^{\dagger}$ and Wolfgang Stephan$^{\S}$\\\mbox{}\\
$^{\dagger}$Theoretical
  Sciences Unit, \\Jawaharlal Nehru Centre for Advanced Scientific Research, \\Jakkur P.O., Bangalore 560064, India  \\
  $^{\S}$ Section of Evolutionary Biology, Department of Biology,
  \\ Ludwig-Maximilians University of Munich, Planegg-Martinsried, Germany}

\maketitle

\newpage

\noindent
Running head: Dynamical response of polygenic traits
\bigskip


\bigskip

\noindent
Correspondence:
\texttt{jain@jncasr.ac.in}, \texttt{stephan@bio.lmu.de}

\bigskip
   
\noindent
\textbf{Abstract:} 
We consider an infinitely large population under stabilising selection
and mutation in which the allelic effects determining a polygenic trait
vary between loci. We obtain analytical expressions for the
stationary genetic variance as a function of the distribution of
effects, mutation 
rate and selection coefficient. We also study the dynamics of the allele
frequencies, focussing on short-term evolution of the phenotypic
mean as it approaches the optimum after an environmental change. We
find that when most effects are small, the genetic variance does not
change appreciably during adaptation, and the time until the
phenotypic mean reaches the optimum is short if the number of loci is
large. However, when most effects are large, the change of the
variance during the adaptive process cannot be neglected. In this
case, the short-term dynamics may be described by \tc{black}{those of a few loci} of large effect. Our results may be used to understand polygenic
selection driving rapid adaptation. 

\newpage

\tc{black}{In the study of fast adaptation, polygenic selection may be more important than selection on single genes. At single genes, strong selection driving fast adaptation generally leads to the rapid fixation of beneficial alleles or at least to huge allele frequency shifts between populations. Classical examples of this type of adaptation are the case of industrial melanism in moths \citep{Hof:2011} and insecticide resistance in Drosophila \citep{Daborn:2002}. However, in many (if not most) cases of fast adaptation, such as in island lizards that are able to adapt very quickly to a changing vegetation (e.g. revealed in an experimental evolution study by \citet{Kolbe:2012}), small shifts of allele frequencies at many loci may be sufficient to move a phenotype toward a new optimum under changed environmental conditions.}

\tc{black}{There is a large and growing body of literature on the detection of adaptive signatures in molecular population genetics. Following pioneering work of \citet{Smith:1974}, the impact of positive selection on neutral DNA variability (Òselective sweepsÓ) has attracted much interest. This theory has been applied to huge datasets that emerge from modern high-throughput sequencing. A large number of statistical tests have been developed to detect sweep signals and estimate the frequency and strength of selection \citep{Kim:2002, Nielsen:2005b, Pavlidis:2010}. However, most theory so far excludes the phenotypic side of the adaptive process (except for fitness). Usually, selection is simply modeled as a constant force that acts on a new allele at a single locus. This is in striking contrast to the classical phenotype-based models of adaptation that are successfully used in quantitative genetics \citep{Barton:2002}. These models typically assume that adaptations are based on allele frequency shifts of small or moderate size at a large number of loci. Also, adaptation does not require new mutations, at least in the short term. Instead, selection uses alleles that are found in the standing genetic variation. Genome-wide data of the past few years show that this quantitative genetic view is relevant. In particular, association studies confirm that quantitative traits are typically highly polygenic. High heritabilities most probably result from standing genetic variation at a large number of loci with small individual effect. Also, local adaptation to environmental clines involves moderate frequency shifts at multiple loci \citep{Hancock:2010}. As a consequence, there is growing evidence that the molecular scenario of sweeps only covers part of the adaptive process and needs to be revised to include polygenic selection.}
	
\tc{black}{As genome-wide association studies (GWAS) yield information about the distribution of single nucleotide polymorphisms (SNPs) relevant to quantitative traits \citep{Visscher:2012}, it is important to understand the models of polygenic selection in terms of the frequency changes of molecular variants, i.e. in terms of population genetics. So far, however, the dynamics of only very simple polygenic models have been studied and applied to data (e.g. \citet{Turchin:2012}). In this article, we analyze the dynamics of a quantitative trait given by a much more general model that was originally proposed by \citet{Wright:1935} and recently re-visited by \citet{Vladar:2014}. These authors consider an infinitely large population evolving under stabilizing selection and mutation. Following the observations from many empirical studies (particularly from biomedicine, see \citet{Visscher:2012}), they assume that the effects are locus-dependent. Furthermore, they consider additivity of the effects and linkage equilibrium between loci.}

Here we study this model and obtain some analytical 
results on the stationary genetic variance and also the dynamics of
the phenotypic mean. 
We show that the stationary genetic variance may exhibit nonmonotonic
dependence on the shape of the distribution of effects. We also study how the
trajectories of the allele frequencies and the mean trait respond to a
sudden environmental shift. \tc{black}{When most effects are small, as
  is the case in experiments on {\it 
  Drosophila} \citep{Mackay:2004}, in livestock
\citep{Hayes:2001,Goddard:2009} and for human height
\citep{Visscher:2008}, a simple analysis} shows
that the magnitude of the deviation of the phenotypic mean from the
optimum decays roughly exponentially with time and approaches zero
over a time scale that is inversely proportional to the initial
genetic variance. When most effects are large, the short-term 
dynamics of the mean and variance can be understood by considering \tc{black}{a few loci with large effects}.

\bigskip
\centerline{MODEL WITH LOCUS-DEPENDENT EFFECTS}
\bigskip

We consider the  $\ell$-locus model recently analysed by
\citet{Vladar:2014} where each locus is biallelic. \tc{black}{The $+$ 
allele at site $i$ has frequency $p_i$ 
while the $-$ allele occurs with frequency $q_i=1-p_i$. The effects
are assumed to be 
additive so that the trait value is $z= \sum_{i=1}^\ell
\textrm{sgn(i)} ~\gamma_i$, where $\textrm{sgn(i)}=\pm 1$ denotes the sign
of the genotypic value of locus $i$ and $\gamma_i > 0$ is the effect
of the allele at the $i$th locus.}   
The loci are assumed to be in linkage equilibrium so that the joint
distribution of effects at the loci factorises. As a result, the $n$th 
cumulant $c_n$ of the phenotypic effect, obtained on averaging over the
population distribution, can be written as the sum over the 
corresponding quantities at individual loci. The first three cumulants
viz. mean $c_1$, 
variance $c_2$ and skewness $c_3$ are given by \citep{Burger:1991} 
\begin{subequations}
\bea
c_1 &=& \sum_{i=1}^\ell \gamma_i (p_i-q_i) \label{meandefn} \\
c_2 &=& 2 \sum_{i=1}^\ell \gamma_i^2 p_i q_i \label{vardefn} \\
c_3 &=&  2 \sum_{i=1}^\ell \gamma_i^3 (q_i-p_i) p_i q_i ~.\label{skewdefn}
\eea
\label{mom}
\end{subequations}

The allele frequency evolves in time under selective pressure and is
given by \citep{Barton:1986} 
\be
\frac{\partial p_i}{\partial t} \approx p_i(t+1)-p_i(t) = \frac{p_i
  q_i}{2 {\bar w}} \frac{\partial {\bar w}}{\partial p_i} ~,
\label{Wright}
\ee
where ${\bar w}$ is the average fitness of the population. For large
$\ell$, as the trait value $z$ of an individual can be treated as a 
continuous variable, from (\ref{meandefn}) and (\ref{vardefn}), we obtain 
\be
{\bar w}=\int_{-\infty}^\infty dz ~p(z)~w(z)=1- \frac{s}{2}
(c_2+(\Delta c_1)^2) \approx e^{-\frac{s}{2} (c_2+(\Delta c_1)^2)} ~,
\label{fitness}
\ee
where the approximate equality sign holds because $s$ is assumed to be
small. In the above expression, 
$w(z)=1-(s/2) (z-z_o)^2$ is the fitness distribution of the phenotypic
trait under stabilising selection, $z_o$ the
phenotypic optimum and $\Delta c_1=c_1-z_o$ the mean deviation from
$z_o$. Thus the maximum fitness
(namely, one) is 
obtained when the population is at the phenotypic optimum and has no
genetic variance. \tc{black}{Inserting equations (\ref{mom}) and (\ref{fitness}) in
(\ref{Wright}) and} accounting for mutations, 
we obtain the following basic equation for the evolution of allele
frequencies, 
\bea
\frac{\partial p_i}{\partial t} = -\frac{s \gamma_i^2}{2} p_i q_i \left(2
\frac{\Delta c_1}{\gamma_i}+q_i-p_i \right)+ \mu (q_i-p_i)
~,~i=1,...,\ell ~,
\label{pevoleqn}  
\eea
where $\mu$ is the probability of (symmetric) mutation between the $+$ and
$-$ allele at locus $i$. Note that the equation (1) of
\citet{Vladar:2014} is obtained by replacing $s$ by $2 s$ in the above
equation.

On the right hand side (RHS) of (\ref{pevoleqn}), the first term (in
the first parenthesis) expressing the mean deviation from the
optimum corresponds to directional selection toward the phenotypic
optimum: if the mean is above (below) the optimum, the allele 
frequencies decrease (increase). However, once the phenotypic mean is
sufficiently 
close to the optimum, stabilising selection (described by
the second term) takes over.

One of the difficulties in solving (\ref{pevoleqn}) is
that it involves the mean $c_1$ which depends on all the allele
frequencies. Moreover, it has been shown that the differential
equations for the cumulants do not close: each one not only involves
two higher cumulants, but also contains terms that can not be written
in terms of other cumulants \citep{Barton:1987,Burger:1991}.  

\bigskip
\centerline{GENETIC VARIANCE IN THE STATIONARY STATE}
\bigskip

In the stationary state in which the left hand side of (\ref{pevoleqn})
vanishes, if the mean $c_1^*= z_o$, the allele 
frequency $p_i^*$ has three solutions, namely  $1/2$ and $(1\pm \sqrt{1-(\hat
  \gamma/\gamma_i})^2)/2$, where ${\hat \gamma}=2 \sqrt{{2 \mu}/{s}}$.  
The latter two solutions are stable for $\gamma_i > {\hat \gamma}$,
and therefore the allele frequency is close to fixation when the
effects are large. For $\gamma_i < {\hat \gamma}$, the effects are
small and the stationary state solution $p_i^*=1/2$ is the only stable
solution for the allele frequency \citep{Vladar:2014}. From these
results, the stationary genetic variance (\ref{vardefn}) is easily seen to be
\citep{Vladar:2014} 
\be
c_2^*= \frac{4 \mu}{s} n_l+ \frac{1}{2} \sum_{\gamma_i < {\hat
    \gamma}} \gamma_i^2 ~,
\ee
where, for large $\ell$, the number of effects larger than ${\hat
  \gamma}$ is given by $n_l= \ell \int_{\hat \gamma}^\infty d \gamma
~p(\gamma)$ with $p(\gamma)$ being the distribution of effects.   
Thus the genetic variance in the stationary
state can be neatly written as  
\bea
c_2^* = \frac{\ell}{2} \left[ {\hat \gamma^2} \int_{\hat
    \gamma}^\infty d \gamma ~p(\gamma)+ \int_{0}^{\hat \gamma} d
  \gamma  ~\gamma^2 ~p(\gamma)\right] ~, \label{c2int} 
\eea
where the first (second) term is the contribution
from loci with large (small) effects.

\tc{black}{The gamma distribution $p(\gamma) \sim \gamma^{k-1} e^{-k \gamma/{\bar
    \gamma}}$ with shape parameter $k > 0$ and mean ${\bar
  \gamma}$ has been employed to fit the distribution of QTL effects
  \citep{Hayes:2001}. This distribution is $\tt{L}$-shaped for $k 
  < 1$ and bell-shaped for $k > 1$, while for $k=1$, it is an
  exponential function. For the gamma distribution, the stationary genetic
  variance (\ref{c2int}) is given by}
\be
c_2^*=\frac{\ell {\hat \gamma}^2}{2}~\left[ \frac{\Gamma(k,k
    \gamma_r)}{\Gamma(k)} + \frac{\Gamma(2+k)-\Gamma(2+k,k
    \gamma_r)}{\gamma_r^2 k^2 \Gamma(k)}\right] ~,
\label{c2avg} 
\ee
where $\gamma_r={\hat \gamma}/{\bar \gamma}$ and
$\Gamma(a,b)=\int_b^\infty dt~t^{a-1} e^{-t}$ is the incomplete gamma
function. For the special case of exponentially distributed effects 
($k=1$), (\ref{c2avg}) simplifies to give $c_2^*={\ell {\bar \gamma}^2}  (1-
e^{-\gamma_r} (1+ \gamma_r))$.

For ${\bar \gamma} \gg {\hat \gamma}$, the House of
Cards (HoC) variance, namely $c_2^*=\ell {\hat 
  \gamma}^2/2 =4 \mu \ell/s$ \citep{Turelli:1984}, is obtained when ${\hat
  \gamma}$ is finite but ${\bar 
  \gamma} \to \infty$ for all $k$. In the opposite case (${\bar \gamma}
\ll {\hat \gamma}$), we have $c_2^*=\ell {\bar \gamma}^2 (k+1)/(2 k)$
which depends on the shape of the distribution of effects. For fixed mean
${\bar \gamma}$, the genetic variance increases monotonically with the
scale ${\hat \gamma}$ since a larger mutation probability increases
the variance. If instead the distribution mean ${\bar \gamma}$ is
increased keeping ${\hat \gamma}$ fixed, the variance increases with
${\bar \gamma}$ toward the HoC value. This is because for fixed $k$,
the width of 
the distribution increases with ${\bar \gamma}$ and therefore larger
effects can be accessed. 

\begin{figure} 
\centering
\subfloat[][]{\includegraphics{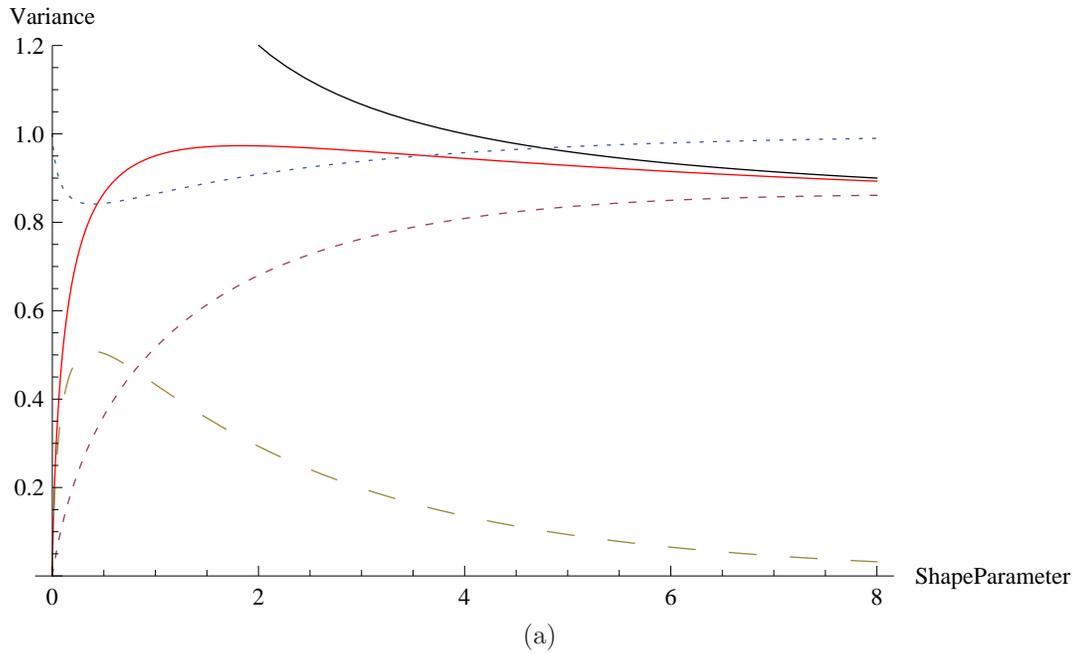}\label{varfig1}}

\subfloat[][]{\includegraphics{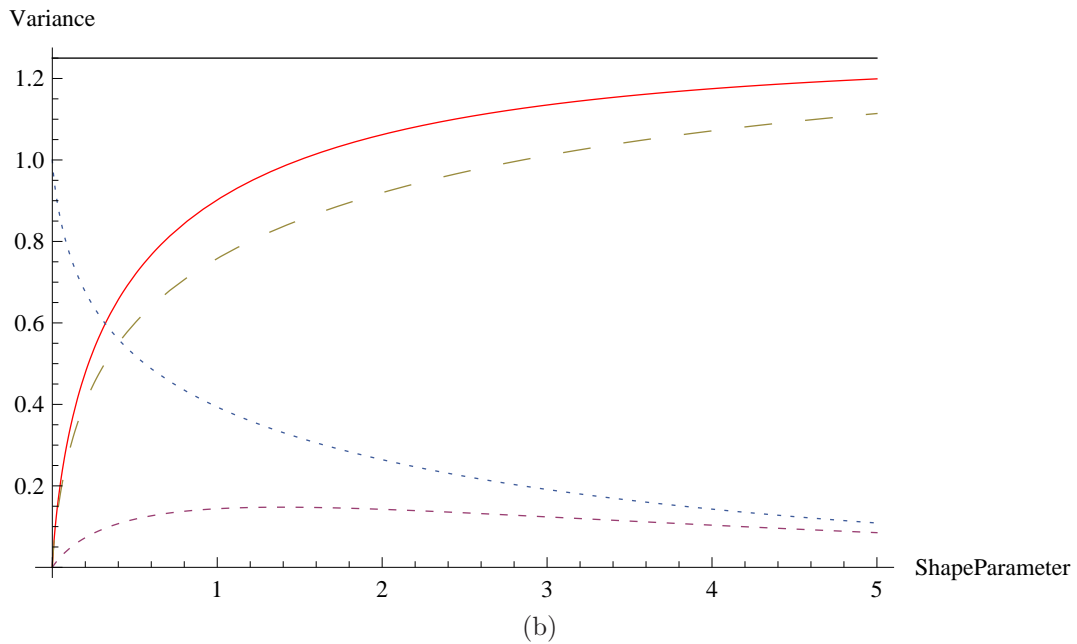}\label{varfig2}}
\caption{Genetic variance in the stationary state as a function of the
  shape parameter $k$ when the effects are distributed according 
  to the gamma function. The plot shows the total genetic variance (solid),
  variance due to small effects (small dashes), large effects
  (large dashes), and the fraction of small effects (dotted) for (a)
  ${\bar \gamma}=0.04, {\hat \gamma}=0.08$ and (b) 
${\bar \gamma}=0.1, {\hat \gamma}=0.05 $ for $\ell=1000$. The  
  asymptotic values $\ell {\bar \gamma}^2 (k+1)/(2 k)$ when ${\hat
    \gamma} > {\bar \gamma}$ and $\ell {\hat \gamma}^2 /2$ when ${\hat
    \gamma} < {\bar \gamma}$ are also shown (top solid curves).} 
\label{varfig}
\end{figure}

The stationary genetic variance has been computed
numerically for various shape parameters when ${\hat \gamma}=0.063,
{\bar \gamma}=0.1$ and $\ell=1000$ in \citet{Vladar:2014}. From 
(\ref{c2avg}), the variance for $k=1, 2, 10$ and $100$ is found to be
$1.32, 1.57, 1.94$ and $2$, respectively, which agrees well with the
numerical data in their Figure 5.  
To understand how the variance depends on the shape parameter, we first
note that with increasing $k$ 
(and fixed $\bar \gamma$), the width of the gamma distribution
decreases. For large $k$, if ${\bar \gamma} > {\hat \gamma}$, the
variance saturates to the HoC variance since almost all loci have
large effects with narrow distributions while in the opposite case,
most effects are small and the variance tends to $\ell {\bar
  \gamma}^2/2$ (see Fig.~\ref{varfig}). For small $k$, irrespective
of whether ${\bar \gamma}$ is above or below ${\hat \gamma}$, we find
that most effects are small. To see this, consider the fraction
$f_s=1-(n_l/\ell)$ of 
loci with small effects which is given by 
\bea
f_s=\frac{(k \gamma_r)^k}{(k-1)!} \int_0^{1} dx~ x^{k-1}
e^{-k \gamma_r x}=1-\frac{\Gamma(k,k \gamma_r)}{\Gamma(k)}  ~.
\label{fsk}  
\eea
If $k < (\gamma_r)^{-1}$, the above equation yields $f_s \approx
{(k \gamma_r)^k}/{k!}$. Then for finite $\gamma_r$ when $k \to 0$, 
we find that $n_s \to \ell$ for any $\gamma_r$, as claimed above. To
summarise, as shown in 
Fig.~\ref{varfig}, for ${\hat \gamma} < {\bar \gamma}$, the variance
increases with $k$ toward the HoC variance, while for ${\hat \gamma} >
{\bar \gamma}$, both $c_2^*$ and $n_s$ are nonmonotonic functions of
$k$.

When the effects are chosen from an exponential distribution,  
the fraction $f_s=1-e^{-\gamma_r}$. On eliminating $\gamma_r$ in favor
of $f_s$ in (\ref{c2avg}) for $k=1$, we  
find the relative contribution of loci with small effects to the
total variance to be  
\be
\frac{c_{2,small}^*}{c_{2}^*}=\frac{2 f_s +(1- f_s) \ln(1 - f_s)
  (2 - \ln(1 - f_s))}{2 f_s + 2 (1- f_s) \ln(1 - f_s)} ~,
\label{relcon}
\ee
which increases as $f_s/3$ for small $f_s$ and approaches unity as
$f_s$ increases toward one. The above expression shows that if $10\%$
of the effects are small, their contribution to the variance is merely
$3\%$, which increases to $21\%$ when $f_s$ is one half. To obtain an equal
contribution from small and large effects, a
disproportionately large fraction ($\sim 83 \%$) of small effects is
required. We are unable to obtain an analytical expression analogous
to (\ref{relcon}) for arbitrary $k$ since $f_s$ is not a simple
function of $\gamma_r$ (see (\ref{fsk}) above). 
However, 
a numerical analysis using (\ref{c2avg}) and (\ref{fsk}) shows that
for the same value of $f_s$, small effects contribute more to the total genetic
variance as the distribution of effects gets narrower. For $f_s=0.1$,
the relative contribution is found to be $2\%, 3\%$ and $5\%$ for
$k=1/2, 1$ and $2$, respectively.  To obtain an equal contribution from loci
with small and large effects, $f_s=0.89, 0.83$ and $0.77$ is needed for the
shape parameters $k=1/2, 1$ and $2$, respectively.

\bigskip
\centerline{DYNAMICS OF THE ALLELE FREQUENCY}
\bigskip

\tc{black}{We now turn to a description of the allele frequency
  dynamics, and will consider the situation when the phenotypic
  optimum is suddenly shifted.} 
As mentioned earlier, due to the term $\Delta c_1$ on the RHS of
(\ref{pevoleqn}), all the frequencies are coupled which 
makes it hard 
to obtain an exact analytical solution of the allele frequency
dynamics. However, under certain conditions, it is
a good approximation to consider only the $c_1$ term in  
(\ref{pevoleqn}) for the initial dynamics and the rest of the terms for
long-term evolution.

To see this, we first note that since $0 < p_i < 1$, the mean
$|c_1(t)| < \sum_i \gamma_i \approx  {\bar \gamma} \ell$. 
For independent and uniformly distributed initial frequencies, 
as the average initial frequency is one half, the leading
  order contribution (in $\ell$) to the initial mean is zero. The
  initial variance is, however, nonzero which gives the
typical initial mean $|c_1(0)| \sim \bar \gamma
\sqrt{\ell}$. 
When the phenotypic optimum $z_o \lesssim {\bar \gamma} {\sqrt \ell}$ and the
number of loci 
is large, the initial value $|\Delta c_1(0)|/\gamma_i \sim \sqrt{\ell}
\gg 1$. Thus at short times, we can neglect $|2 p_i-1|$ (which is
bounded above by one) and the mutation term in comparison to the term
$2 \Delta c_1/\gamma_i$ in (\ref{pevoleqn}). At large enough crossover
time $t_\times$, as explained below, the mean deviation is close to 
zero and the reverse condition holds; {\it i.e.} $2 |\Delta
c_1(t)|/\gamma_i \ll |2 p_i-1|$ in (\ref{pevoleqn}), and we may set
$\Delta c_1 \approx 0$ for later evolution. Biologically these
considerations mean that initially the effects are weaker than the
mean trait deviation, but as the population adapts due to directional
selection, the deviation of the mean from the phenotypic optimum 
becomes smaller than the effects.  

The above argument applies not only to uniformly distributed initial
frequencies but in more general settings as well where
  $|c_1(0)| \sim {\bar \gamma} \ell$ by replacing ${\sqrt{\ell}}$ by
  $\ell$. Here we will focus
on the dynamics of the allele frequency when the optimum is suddenly
shifted to a 
new value $z_f (< \ell {\bar \gamma})$, starting from the population which
is equilibrated to 
a phenotypic optimum value $z_o$. In this situation, as the initial
frequency is close to one half when $\gamma_i < {\hat \gamma}$, the
frequency $|2 p_i(0)-1|$ is obviously negligible compared to $\Delta
c_1(0)/\gamma_i$, whereas for $\gamma_i >  {\hat \gamma}$, $|2
p_i(0)-1|$ is close to one since the initial frequency is close to
either zero or one \citep{Vladar:2014}. 


\noindent{\bf When most effects are small:} The effects at most of
the loci can be smaller than the scale ${\hat \gamma}$ either if $k$ is large
and ${\bar \gamma} < {\hat \gamma}$, or if $k$ is small. Then for most
loci, at short times, the full model
defined by (\ref{pevoleqn}) can be approximated by  
\begin{subequations}
\bea
\frac{\partial p_i}{\partial t} &=&  -s \gamma_i p_i q_i \Delta
c_1 \label{pevolshort} \\ 
\frac{\partial c_n}{\partial t}  &=& -s  \Delta c_1 c_{n+1} ~,~
n \geq 1 ~, \label{cumshort}
\eea
\end{subequations}
where the last equation for cumulants is obtained from the results of
\citet{Burger:1991}. Equation (\ref{cumshort}) for $n=1$ shows that
the magnitude of the mean deviation decreases with time, and for the 
phenotypic optimum smaller than the maximum attainable value of the mean
($z_o \ll {\bar \gamma} \ell$), 
the trait mean becomes close to the phenotypic optimum at large times
(see Fig.~\ref{smallres_cum}a).   
We now assume that the variance $c_2$ is independent of time and stays at
its initial value $c_2(0)$ \tc{black}{\citep{Chevin:2008}}. As explained in Appendix~\ref{Appcvar},
this approximation is good when a combination of the initial cumulants
is small (see also Fig.~\ref{smallres_cum}b). This allows us to solve
(\ref{pevolshort}) and (\ref{cumshort}), and we immediately find that 
\bea
\Delta c_1(t) &=& \Delta c_1(0) e^{-c_2(0) s t } \label{CHsoln2}\\
p_i(t) &=& \frac{p_i(0)}{p_i(0)+q_i(0) e^{\frac{\gamma_i \Delta
      c_1(0)}{c_2(0)} (1-e^{-c_2(0) s t })}} ~. \label{CHsoln}
\eea
Equation (\ref{CHsoln2}) shows that the mean deviation approaches zero
over a time scale $t_\times \sim (s c_2(0))^{-1}$. 

Next we analyse the long-term evolution of the allele frequencies. As
Fig.~\ref{smallres_cum}a shows, there is a small but nonzero mean 
deviation $\Delta c_1^*$ in the stationary state. Taking this into 
consideration and accounting for the other terms in (\ref{pevoleqn}), for
$t >t_\times$, we can write    
\be
 \frac{\partial p_i}{\partial t}=-\frac{s}{2} \gamma_i^2  p_i q_i (1-2 p_i+ \frac{2 \Delta c_1^*}{\gamma_i})+ \mu (1-2 p_i)~.
  \label{pevollong}
 \ee
For $\Delta c_1^*=0$, the above equation can be easily solved to give
\be
p_i^{(\pm)}(t)= \frac{1}{2} \left(1 \pm \sqrt{\frac{1-m_i}{1-M_i(t)}}
\right) ~,~t >t_\times ~,
\label{longtsoln}
\ee
where 
\be
M_i(t)=\frac{4 p_i^2(t_\times)-4 p_i(t_\times)+ m_i}{(2
  p_i(t_\times)-1)^2} ~e^{-\frac{s 
    \gamma_i^2 (1-m_i) (t-t_\times)}{2}} ~,~t > t_\times ~,
\label{Meqn}
\ee
$m_i=({{\hat \gamma}}/{\gamma_i})^2$ and $p_i(t_\times)$ is obtained
from (\ref{CHsoln}).  We check that the stationary state solutions
$(1\pm \sqrt{1-m_i})/2$ and $1/2$ are obtained from the above result
for $m_i < 1$ and $> 1$, respectively. Furthermore, the solution
$p_i^{(+)}(t)$ is obtained for $p_i(t_\times) 
> 1/2$ and $p_i^{(-)}(t)$ for $p_i(t_\times)< 1/2$.   

Figures~\ref{smallres_cum} and \ref{smallres_freq} show a comparison
between the numerical solution of (\ref{pevoleqn}) and the
approximation described above, when the initial condition is the
stationary state of the population equilibrated to a phenotypic
optimum $z_o$. The initial mean deviation 
$\Delta c_1(0)$ is seen to be close to $-z_f$, and the initial variance
$c_2(0)=c_2^*$ for the zero mean deviation is $0.0967$, which is
close to the value $0.131$ obtained from the set of  effects used 
in Figs.~\ref{smallres_cum} and \ref{smallres_freq}. 
As Fig.~\ref{smallres_cum} shows, the dynamics of the mean deviation
are captured well by (\ref{CHsoln2}) and 
approach a stationary value close to zero ($\Delta c_1^*\approx
-0.016$) in about $1500$ generations. The variance also 
evolves with time, but the change is not substantial and the
approximation $c_2(t) \approx c_2(0)$ is good over the time scale 
directional selection toward the phenotypic optimum operates.  
Equation (\ref{CHsoln2}) also indicates that directional selection
toward the optimum will occur faster when the initial variance is
large since $t_\times \sim 1/c_2^*$. 

Since the stationary genetic
variance displays a nonmonotonic dependence on the shape parameter $k$
of the gamma distribution (see Fig.~\ref{varfig}), the relaxation time for
the mean deviation 
is expected to decrease and then increase with increasing $k$. Indeed, 
as the inset of Fig.~\ref{smallres_cum}a shows, the difference
$c_1(t)-c_1^*$ (which, by definition, is zero in the equilibrium state
for all $k$) equals a reference value $-0.05$ at time $360, 340$ and $370$
for $k=1, 5$ and $20$, respectively. 

The allele frequency dynamics are shown in
Fig.~\ref{smallres_freq}. We see that while the short-term
dynamics can be accurately described by (\ref{CHsoln}) for loci with
effects smaller than or close to the distribution mean, there is a substantial
difference when the effects are larger than the mean. \tc{black}{This is
  because for such loci, the initial frequency is not close to half and the
  term involving $q_i-p_i$ on the RHS of (\ref{pevoleqn}) can not be
  neglected.} For $t > 
t_\times$, the long-term behavior described by (\ref{pevollong}) is
shown with $\Delta c_1^*=0$ and the actual mean deviation.  


\noindent{\bf When most effects are large:} When 
${\bar \gamma} > {\hat \gamma}$ and $k$ is large, the number of loci
with large effects is also large, and the initial allele frequencies
are close to either zero or one. In this parameter regime,  
both the variance and the skewness may change appreciably during
directional selection toward the optimum, and the constant-variance
approximation discussed above is not suitable. 
However, at very short times when $\Delta c_1(t)$ is close to its
initial value, the solution (\ref{CHsoln}) for the allele frequency gives 
\be
p_i(t) \approx \frac{1}{1+\frac{q_i(0)}{p_i(0)} e^{\gamma_i
      \Delta c_1(0) s t}} ~.
\label{largeeff1}
\ee
From the above equation, we first note that the allele frequency 
at large-effect loci changes fast as expected intuitively. \tc{black}{Equation
(\ref{largeeff1}) also shows that for $\Delta 
c_1(0) < 0$, the allele frequency quickly increases toward unity, if
the initial frequency is close  
to unity and therefore does not contribute to the dynamics of
the variance or skewness. Thus to understand the short-term dynamics,
we need to focus our  attention on large-effect loci with  
low initial allele frequency for negative initial
mean deviation. Similar remarks apply to the situation when $\Delta 
c_1(0)$ is positive where the large-effect loci with high
initial frequency determine the dynamics.}

In the following, we assume that $\Delta c_1(0) < 0$ and consider the time evolution
of the allele frequency $P$ of 
the largest effect locus with lowest initial frequency.  Figure~\ref{largeres1} shows that the
allele frequency $P$ sweeps to fixation, but the 
frequency of the next relevant locus (i.e. the 
next largest effect locus with low initial frequency) does not. In such a case, we can approximate the mean $c_1$ and variance $c_2$ by the contribution from the frequency $P$ with effect $\Gamma$, and obtain
\begin{subequations}
\bea
c_1(t) & \approx & 2 \Gamma (P-P_0)+c_1(0) \label{largePcum1} \\
c_2(t) & \approx & 2 \Gamma^2 (P Q-P_0 Q_0) +c_2(0) \label{largePcum2} ~,
\eea
\end{subequations}
where $P_0 \equiv P(0)$. Then using the above expression for the phenotypic 
mean in (\ref{pevoleqn}) and neglecting mutations (since most effects
are large), we get  
\be
\frac{\partial P}{\partial t} = -s \Gamma^2 P (1-P) (P +  \alpha) ~,
\label{largePdiff}
\ee
where
\tc{black}{ 
\be
\alpha= \frac{\Gamma+2 \Delta c_1(0)}{2 \Gamma}-2 P_0 ~.
\ee}
We thus find that the allele frequency $P$ is a solution of the
following equation : 
\be
\frac{(P/P_0)^{1+\alpha}}{(Q/Q_0)^\alpha} = e^{-s \Gamma^2 \alpha
  (1+\alpha) t} \frac{P+\alpha}{P_0+\alpha} ~.
\label{largeP}
\ee
An explicit solution of (\ref{largeP}) seems hard to obtain since
$\alpha$ is in general not an integer. However, for large and negative
$\alpha$, the above equation yields 
\be
P = \frac{1}{1+\frac{Q_0}{P_0} e^{-s \Gamma^2 |\alpha| t}} ~.
\label{largePapp}
\ee
Thus for $z_f \gg \Gamma$, the frequency $P$ sweeps to fixation in a
time of order $(s \Gamma z_f)^{-1}$.  

Figure~\ref{largeres1} shows the allele frequency of the largest
effect locus with lowest initial frequency obtained using
(\ref{pevoleqn}). It agrees reasonably well with the solution of
(\ref{largeP}) and the expression (\ref{largePapp}) where
$\alpha=-1.43$.  
In Fig.~\ref{largeres2}, the dynamics of the first two cumulants given
by (\ref{meandefn}) and (\ref{vardefn})  are compared with the
approximate expressions (\ref{largePcum1}) and (\ref{largePcum2}),
respectively, and we see a good agreement.  

\tc{black}{A detailed numerical analysis of the set of parameter values
  of  Fig.~\ref{largeres1} suggests that the dynamics of this example
  can be understood by considering one, two or three of the
  largest-effect loci. When only the largest-effect locus was
  required, the second largest effect was much lower than the largest
  one.}

\bigskip
\centerline{DISCUSSION}
\bigskip

One of the fundamental questions in adaptation is whether the adaptive
process is governed by many loci of small effect or few loci of 
large effect \citep{Orr:2005}. However, which effects are small, and
which large? 
\citet{Vladar:2014} have provided a scale ${\hat \gamma} \sim 
\sqrt{\mu/s}$ for the size of effects, which is a function
of basic population genetic parameters, namely mutation probability
$\mu$ and selection coefficient $s$, relative to which an effect is 
defined as large or small. For a given distribution of effects, assumed
here to be a gamma distribution with mean ${\bar \gamma}$ and shape
parameter $k$, an effect is small
(large) if it is below (above) ${\hat \gamma}$. But for fixed ${\bar
  \gamma}$ and ${\hat \gamma}$, whether most or a few effects are
small depends on the shape parameter $k$: 
for small $k$, most effects are small, but for large $k$, the number
of small effects depends on the ratio $\gamma_r={\hat \gamma}/{\bar \gamma}$.

\noindent{\bf Genetic variance in stationary state:} Here we have
provided analytical expressions for the stationary genetic variance
$c_2^*$ when the effects are locus-dependent. We find that when most
effects are small, $c_2^*$ is a 
nonmonotonic function of the shape parameter of the gamma
distribution (going through a maximum for intermediate values of $k$;
see Fig.~\ref{varfig}). In contrast, it increases monotonically when most 
effects are 
large. As Fig.~\ref{varfig} shows, when the
shape distribution is narrow, large (small) effects contribute most to
the variance when $\gamma_r > 1 (< 1)$. However for
broad distributions, although the number of small effects is large,
small effects do not contribute much when $\gamma_r < 1$. The HoC variance
is obtained irrespective of $k$ when $\gamma_r \to 0$ since all 
loci have large effect in this limit. As noted previously 
\citep{Vladar:2014}, HoC provides an upper bound on the genetic variance. 

\noindent{\bf Dynamics when most effects are small:} \tc{black}{As the  
distribution of QTLs  measured in 
experimental and natural populations \citep{Mackay:2004,Hayes:2001,Goddard:2009,Visscher:2008} find most effects to be small},  
it is 
important to study this situation in detail. Here we have obtained
analytical expressions for the dynamics by assuming the genetic 
variance to be constant. Although the fact
that the variance does not change much in time when most effects are small
was observed numerically in \citet{Vladar:2014}, an explanation of
this behavior was not provided. Here, as explained in
Appendix~\ref{Appcvar}, it is a good approximation to assume the variance to be
time-independent provided  
the product of the initial values of the mean deviation and skewness
is small. 

In the absence of mutations, \citet{Chevin:2008} have 
considered the effect of background with a time-independent genetic
variance on the frequency at a single focal locus. Their results
match the ones 
obtained here using the short-term dynamics model with directional
selection only; in particular, 
(\ref{CHsoln2}) and (\ref{CHsoln}) match the results (21)
and (25) of \citet{Chevin:2008}, respectively, on identifying their parameters
$\omega^2$ and $a$ with $1/s$ and $\gamma$ from this study.

Our basic result concerning the dynamics of
the phenotypic mean is that it relaxes over a time scale which is 
inversely proportional to the initial variance. Since the variance
is of order $\ell$, we thus have the important result that the mean
approaches the optimum faster if a larger 
number of loci is involved. Moreover, this time depends
nonmonotonically on the shape parameter of the gamma distribution. 
Note that the phenotypic mean deviation relaxes to zero when the
phenotypic optimum is far below the upper 
bound $\bar \gamma \ell$ on the phenotypic mean. However, when the phenotypic 
optimum exceeds the maximum typical value of the mean, such that the
mean deviation remains substantially different from zero 
at late times, (\ref{cumshort}) shows that all higher cumulants vanish
at the end of the phase of directional selection.

\noindent{\bf Dynamics when most effects are large:} 
When the initial mean deviation is moderately large (and negative),
the genetic variance changes by a large amount over the time scale
directional selection occurs and the dynamics can be understood by
considering \tc{black}{a few loci} whose effect is large but initial frequency
is low. 

However, for larger mean deviations (but smaller than $\ell
{\bar \gamma}$), \tc{black}{a few large-effect loci do not} completely
capture the dynamics of the mean and the variance. As
Fig.~\ref{largeresA} shows, the initial increase of the absolute
mean deviation and the transient rise of the variance can be explained
by considering the large-effect locus. At later times, however, as the change
in variance is small, we can use the constant-variance approximation
to understand the dynamics of the phenotypic mean deviation until it
nearly vanishes. The constant-variance approximation can also be used
when the initial mean deviation is sufficiently small (see
Fig.~\ref{largeresS}).  

\noindent{\bf Applications:} The approximations presented here hold
for the short-term evolution of phenotypic traits and allele
frequencies. This means that our results may be used to understand
polygenic selection driving rapid adaptation. In this respect, our
most important result is that the mean of a phenotypic trait may
respond faster to a sudden environmental change when the number of
loci is large and most effects are small.  

Evidence for rapid phenotypic evolution has been reported in recent
years from several groups of organisms. For instance in {\it
    Drosophila subobscura}, latitudinal clines of wing size have been
formed within $20$ 
years since this species colonized America
\citep{Huey:2000}. Similarly, in field experiments in which lizard
populations were newly established on small islands in the Bahamas,
the hindlimbs  
adapted very quickly to the different vegetations on the islands 
\citep{Kolbe:2012}. To our knowledge, however, data from GWAS are not
yet available in these cases.  

The theory presented here can also be applied to the large amounts of
GWAS data that have been gathered in model species such as humans and
Drosophila. To analyse the observed allele frequency shifts in SNPs
associated with quantitative traits, such as human height 
\citep{Turchin:2012} and cold tolerance in Drosophila \citep{Huang:2012}, the
results derived in this study provide a more general theoretical basis
than the dynamical equations used in previous analyses
(e.g. \citet{Turchin:2012}).

\noindent{\bf Open questions:} The analytical calculations in this
article work when the phenotypic mean at the equilibrium coincides
exactly  with the 
optimum. However, in the stationary state, there is a small but nonzero mean 
deviation due to which 
the long-term dynamics are not accurately captured, \tc{black}{
  especially when effects are small, as shown in
  Fig.~\ref{smallres_freq}b.} An 
improved calculation of the dynamics that takes a nonzero mean deviation
into account is certainly of interest, for instance to estimate the
frequency of selective fixations (leading to selective sweeps) in this
model \citep{Chevin:2008,Pavlidis:2012,Wollstein:2014}. 

Another open question concerns the generality of our results presented
here. The current model perhaps oversimplifies biological reality in
that it neglects genetic drift and assumes additive effects, symmetric
mutations and free recombination between loci. It can be shown that
the current model (neglecting mutation) can be derived from the
classical symmetric viability model with arbitrary position of the
optimum under the quasi-linkage equilibrium assumption. In the latter
model the probability of selective fixation has been studied
numerically for up to eight loci
\citep{Pavlidis:2012,Wollstein:2014}. However, at present we are
lacking an analytical 
understanding of the role of recombination in this model and how it
relates to the high-recombination limit represented by our current
model. \\

\noindent{\bf Acknowledgements:} We thank two anonymous referees for insightful comments on a previous version of this manuscript and A. Wollstein for technical
support. KJ was supported by a 
fellowship from the Center of Advanced Studies of the LMU Munich. WS
was funded by grants 325/14-1 (Priority Program 1590) and 325/17-1 (Priority Program 1819) of the Deutsche Forschungsgemeinschaft.

\clearpage

\begin{center}
{\bf {\large APPENDIX}}
\end{center}

\appendix
\numberwithin{equation}{section}

\section{Validity of constant-variance approximation}
\label{Appcvar}

\begin{figure} 
\centering
\subfloat[][]{\includegraphics[width=0.7\textwidth,angle=0]{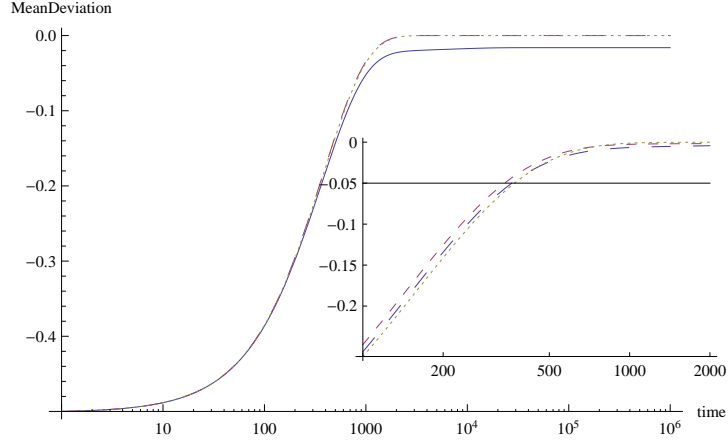}\label{smallres_cum1}}

\subfloat[][]{\includegraphics[width=0.7\textwidth,angle=0]{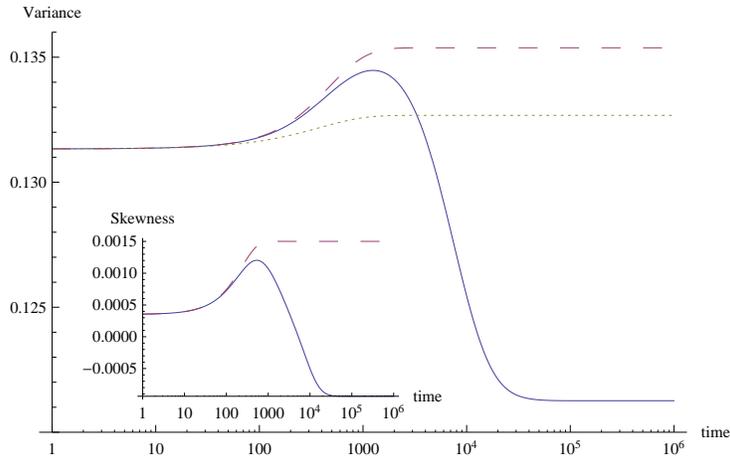}\label{smallres_cum2}}
\caption{Response to change in optimum when most effects are small. 
  The plot shows the results for (a) mean deviation $\Delta c_1(t)$ and
  (b) variance $c_2(t)$ and 
  skewness $c_3(t)$ obtained using the exact numerical solution of the
  full model 
  (solid) and the short-term dynamics model (large dashes). The dotted
  curves show the time-dependent solution (\ref{CHsoln2}) for mean and
  (\ref{valid}) for variance. The parameters are $\ell= 50, s = 0.02,
  \mu=5 \times 10^{-5}, {\hat \gamma}=0.14 > {\bar \gamma}=0.05,
  z_o=-0.0012, z_f=0.5, n_l=5$. The effects are chosen from an
  exponential distribution, and the parameter ${\cal C}=-0.01$ (see 
  Appendix~\ref{Appcvar}). The inset in the top figure shows the
  difference $\Delta c_1(t)-\Delta c_1^*$ as a function of time for 
   the full model when the effects are gamma-distributed
  with shape parameter $k=1$ (large dashes), $5$ (small dashes) and $20$
  (dotted).} 
\label{smallres_cum}
 \end{figure}

\begin{figure} 
\centering
\subfloat[][]{\includegraphics{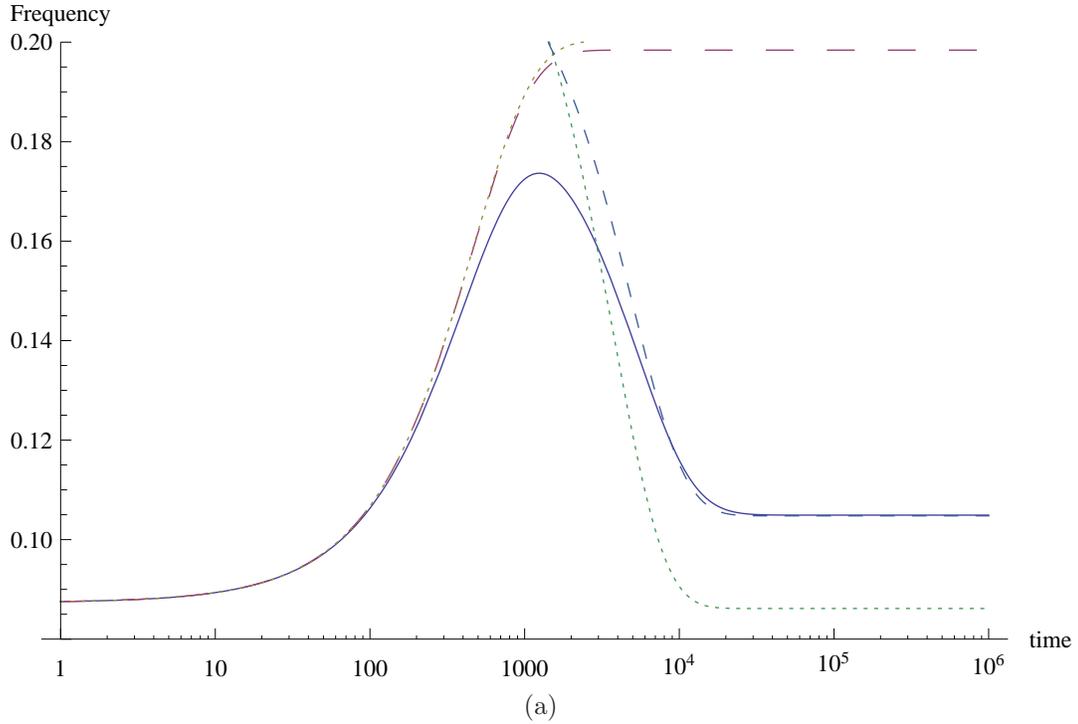}\label{smallres_freq1}}

\subfloat[][]{\includegraphics{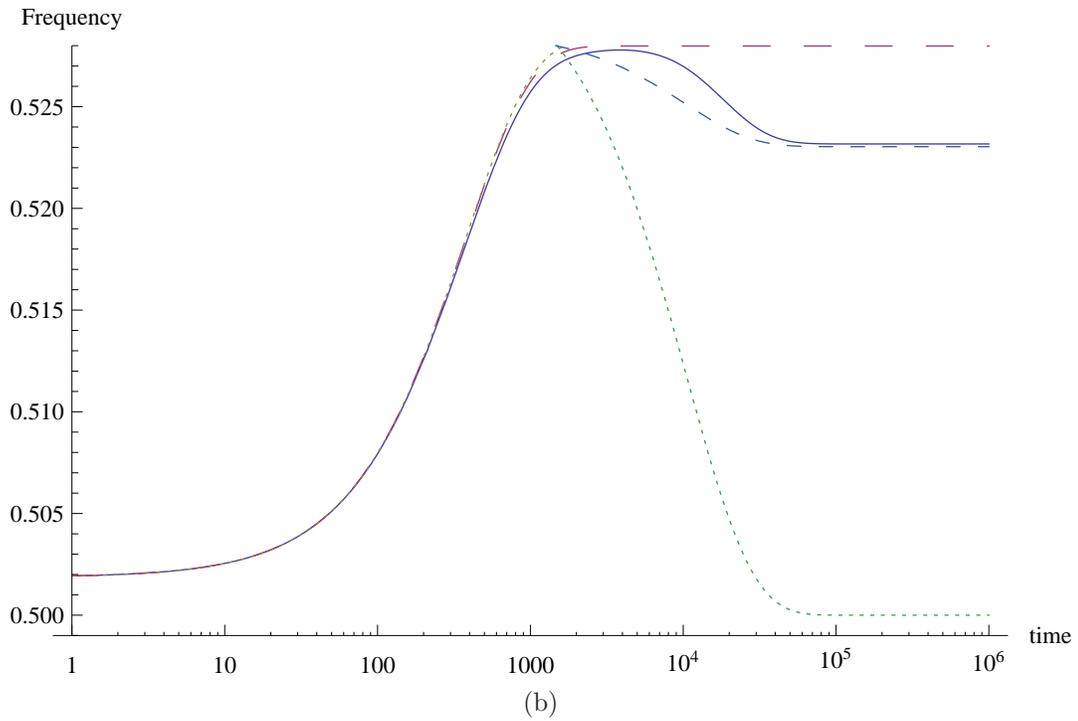}\label{smallres_freq2}}
\caption{Response to change in optimum when most effects are
  small. The plot shows the allele frequencies for two 
  representative loci with (a) $\gamma_i=0.252$ and (b)
  $\gamma_i=0.028$ for the full
  model (solid) and short-term dynamics model (large dashes). The
  dotted curves show the time-dependent solution (\ref{CHsoln}) for $t
  < t_\times$ and (\ref{longtsoln}) for $t > t_\times$ where
  $t_\times=1500$. The dashed curve for $t > t_\times$ is the solution
  of (\ref{pevollong}) with $\Delta c_1^*=-0.016$. The other
  parameter values are the same as in Fig.~\ref{smallres_cum}.} 
\label{smallres_freq}
 \end{figure}

\begin{figure} 
\begin{center}
\includegraphics[width=0.8\textwidth,angle=0]{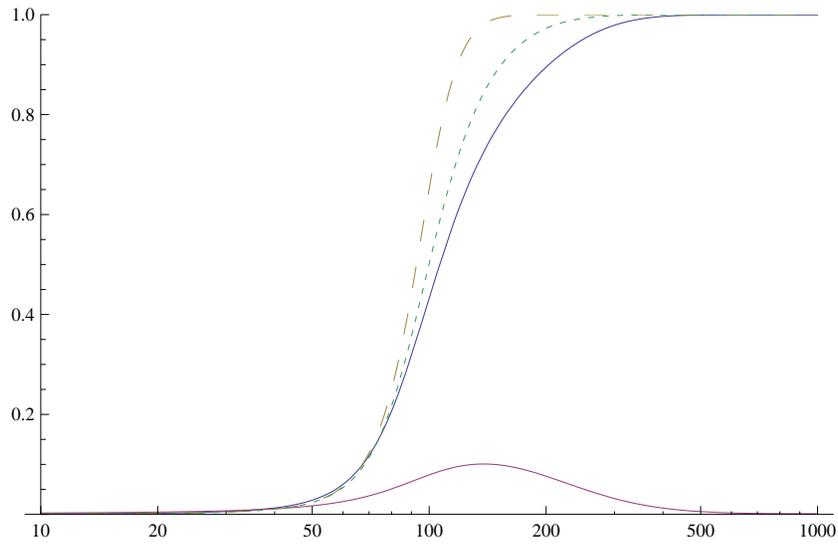}
\end{center}
\caption{Response to change in optimum when most effects are large.  
 The plot shows the exact numerical solution of the full model (solid)
 and the equations (\ref{largeP})  (large dashes) and
 (\ref{largePapp}) (small dashes) for the dynamics 
of the allele frequency $P$ with the largest 
  effect and lowest initial  
frequency ($\Gamma=0.776, P_0=3.3 \times 10^{-4}$). The solid curve at the bottom shows the numerical solution of the full model for the frequency of the next relevant locus with effect size $0.319$ and initial frequency $1.9 \times 10^{-3}$. The parameters are $\ell=
20, s = 0.1, 
\mu=10^{-5}, {\hat \gamma} \approx 0.028 \ll {\bar \gamma}=0.2,
z_o=7.8 \times 10^{-5}, z_f=1.5, n_l=19$. }
\label{largeres1}
 \end{figure}

\begin{figure} 
\centering
\subfloat[][]{\includegraphics{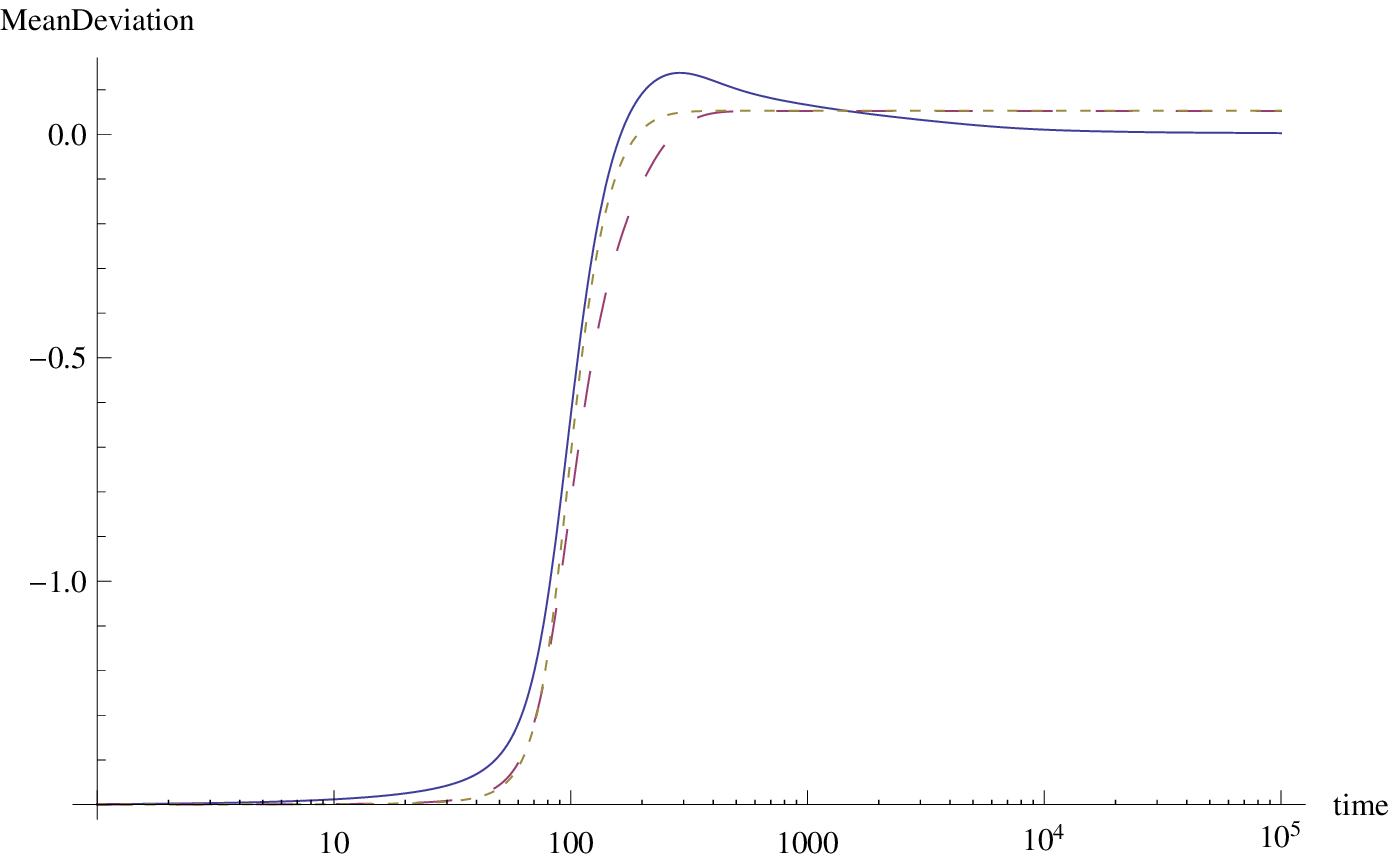}\label{largeres21}}

\subfloat[][]{\includegraphics{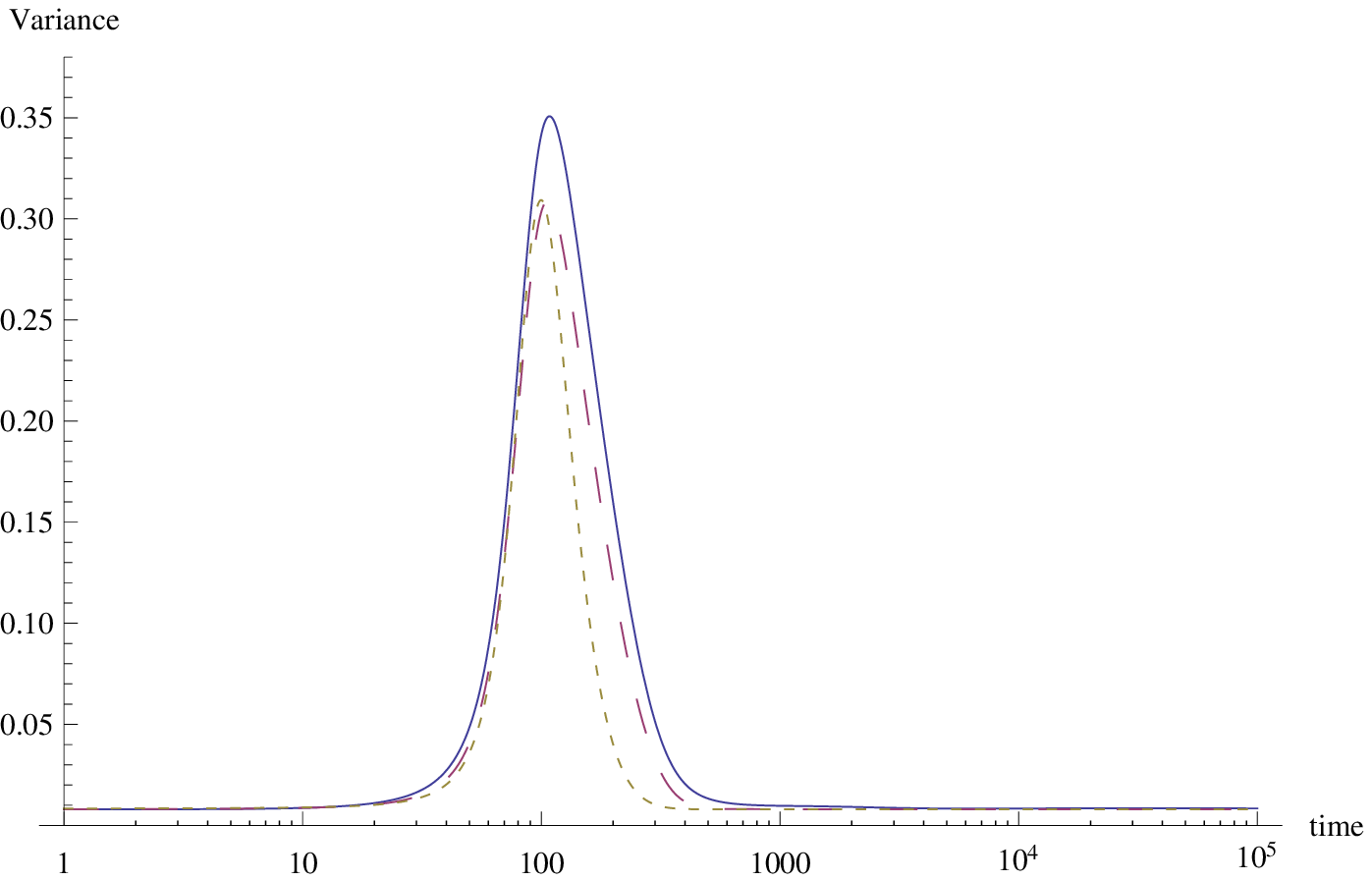}\label{largeres22}}
\caption{Response to change in optimum when most effects are large. 
  Solid lines show the mean deviation (\ref{meandefn}) and variance
  (\ref{vardefn}), while the large dashed curves show  the
  contribution to these cumulants from the locus with the largest
  effect and lowest initial 
frequency ($\Gamma=0.776, P_0=3.3 \times 10^{-4}$). In both cases, the
exact 
numerical solution of the full model is used. The numerical solution of 
(\ref{largePdiff}) (small dashes) is also shown. The other parameter
values are the same as in Fig.~\ref{largeres1}.} 
\label{largeres2}
 \end{figure}

As explained in the main text, we obtain (\ref{CHsoln2}) when the variance
is assumed to be constant in time. To see under what conditions this
approximation holds, we find a correction to the variance by assuming
that the skewness $c_3$ is nonzero and time-independent (see the
inset of Fig.~\ref{smallres_cum}b). Using
(\ref{CHsoln2}) on the RHS of (\ref{cumshort}) for $n=2$, we
immediately get   
\be
c_2(t)=c_2(0) ~\left[1-{\cal C} (1-e^{-c_2(0) s t}) \right] ~,
\label{valid}
\ee
where ${\cal C}=\Delta c_1(0) c_3(0)/c_2^2(0)$. Plugging the above
solution into (\ref{cumshort}) for $n=1$ gives the  
mean deviation as 
\be
\Delta c_1(t)= \Delta c_1(0) ~\textrm{exp}\left[-s t c_2(0) (1-{\cal
    C}) -{\cal C} (1-e^{-s t c_2(0)}) \right] ~.
\ee
The solution (\ref{CHsoln2}) is recovered if the constant ${\cal C}$,
which depends on the initial value of the first three cumulants, is
negligible. 

If we start with the initial condition in which the population is
equilibrated to an optimum and most effects are small, since most
initial allele frequencies are close to one half, the initial variance is
substantial and the skewness is close to zero. In this case, the
constant-variance approximation is expected to 
work well. However, if most effects are large, since most initial
allele frequencies 
are close to fixation, although the skewness remains small, the variance also
becomes small, thus leading to an increase in ${\cal C}$. Then for sufficiently
small mean 
deviations, we may also employ the constant-variance approximation when most
effects are large.



\clearpage

\clearpage

 \clearpage
 
 \clearpage

\clearpage

\clearpage

\clearpage


\begin{center}

{\huge{\bf File S1}}

\hspace{7in}

{\huge Response of polygenic traits under stabilising selection and mutation when loci have unequal effects}

\hspace{1in}

{\huge Supporting Information} 

\hspace{2in}

{\large Kavita Jain and Wolfgang Stephan}
\end{center}

\setcounter{figure}{0}
\setcounter{page}{1}
\makeatletter 
\renewcommand{\thefigure}{S\@arabic\c@figure} 

\rfoot{\thepage}
\lfoot{K. Jain and W. Stephan}

\begin{figure} 
\begin{center}  
\includegraphics[width=0.8\textwidth,angle=0]{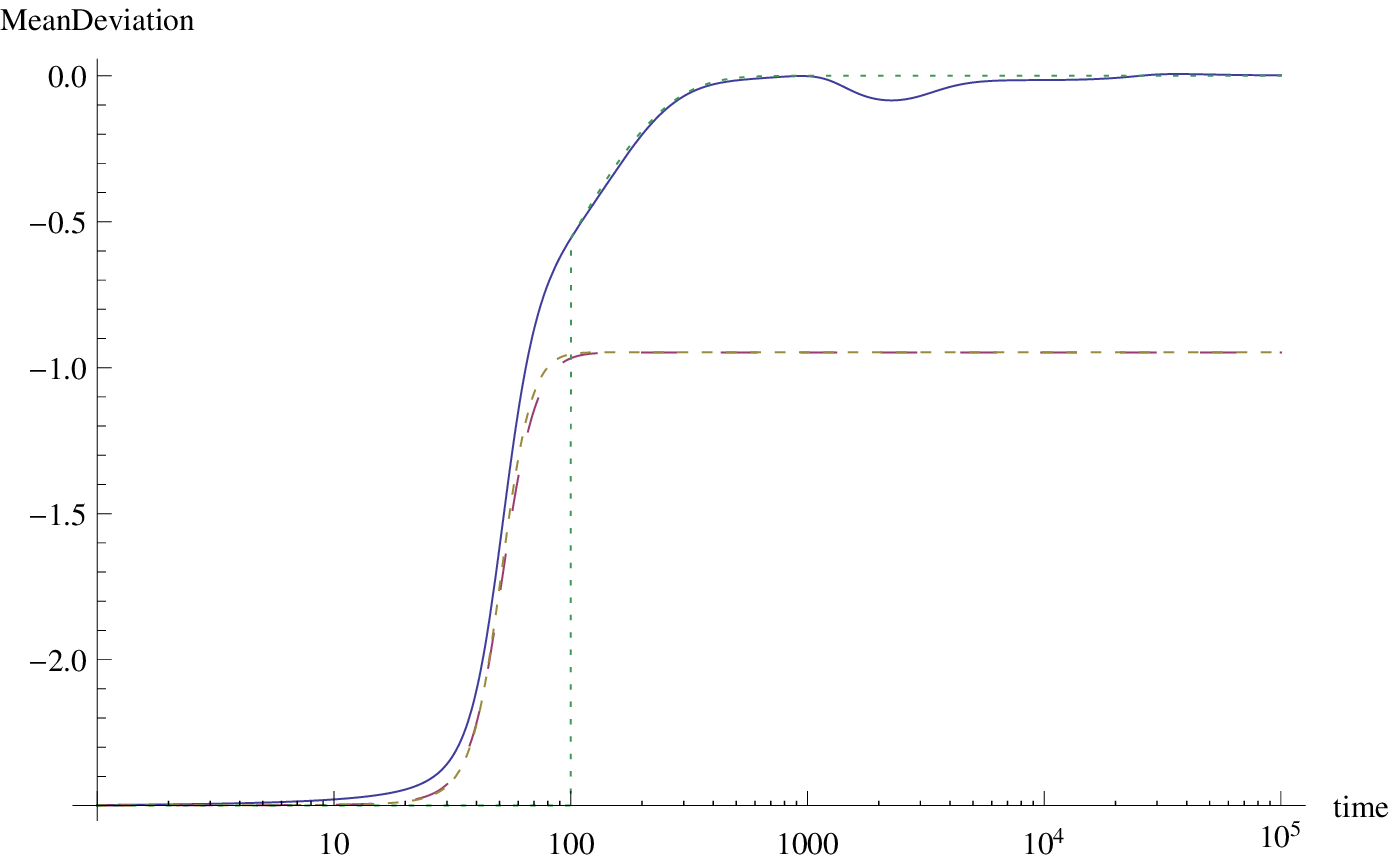}
\includegraphics[width=0.8\textwidth,angle=0]{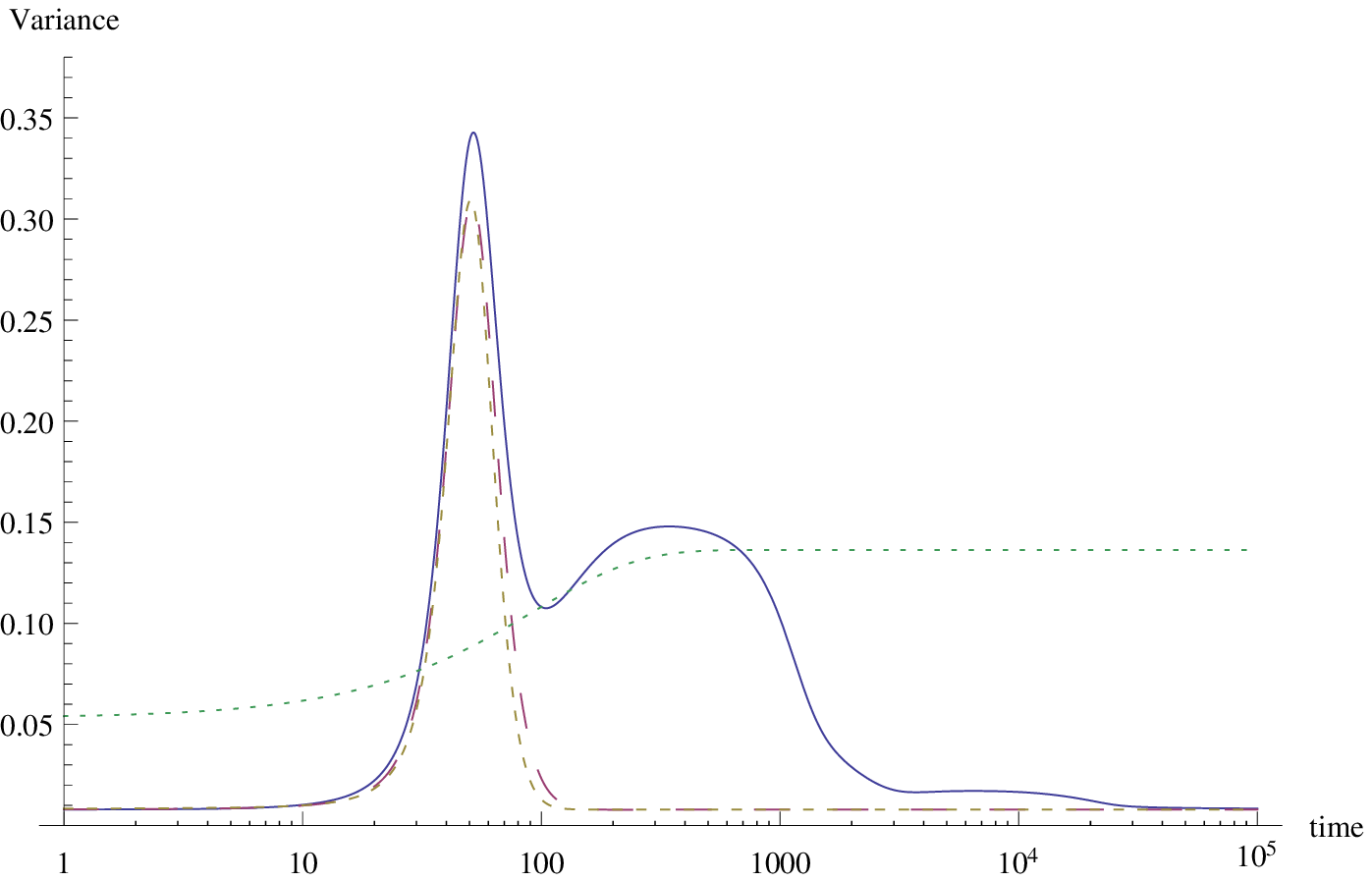}
 \end{center}
\caption{Response to change in optimum when most effects are
  large. Solid lines show the mean deviation (\ref{meandefn}) and
  variance 
  (\ref{vardefn}), while the large dashed curves show  the
  contribution to these cumulants from the locus with the largest
  effect and lowest initial 
frequency ($\Gamma=0.776, P_0=3.3 \times 10^{-4}$). In both cases, the 
exact numerical solution of the full model is used. The numerical solution of 
(\ref{largePdiff}) (small dashes) is also shown. The dotted curves 
show (\ref{CHsoln2}) and (\ref{valid}) for $t > 100$. The final
optimum value $z_f=2.5$ and the other parameter values are the same as in
Fig.~\ref{largeres1}.} 
\label{largeresA}
 \end{figure}
 
 \clearpage
 
 \begin{figure} 
\begin{center}  
\includegraphics[width=0.8\textwidth,angle=0]{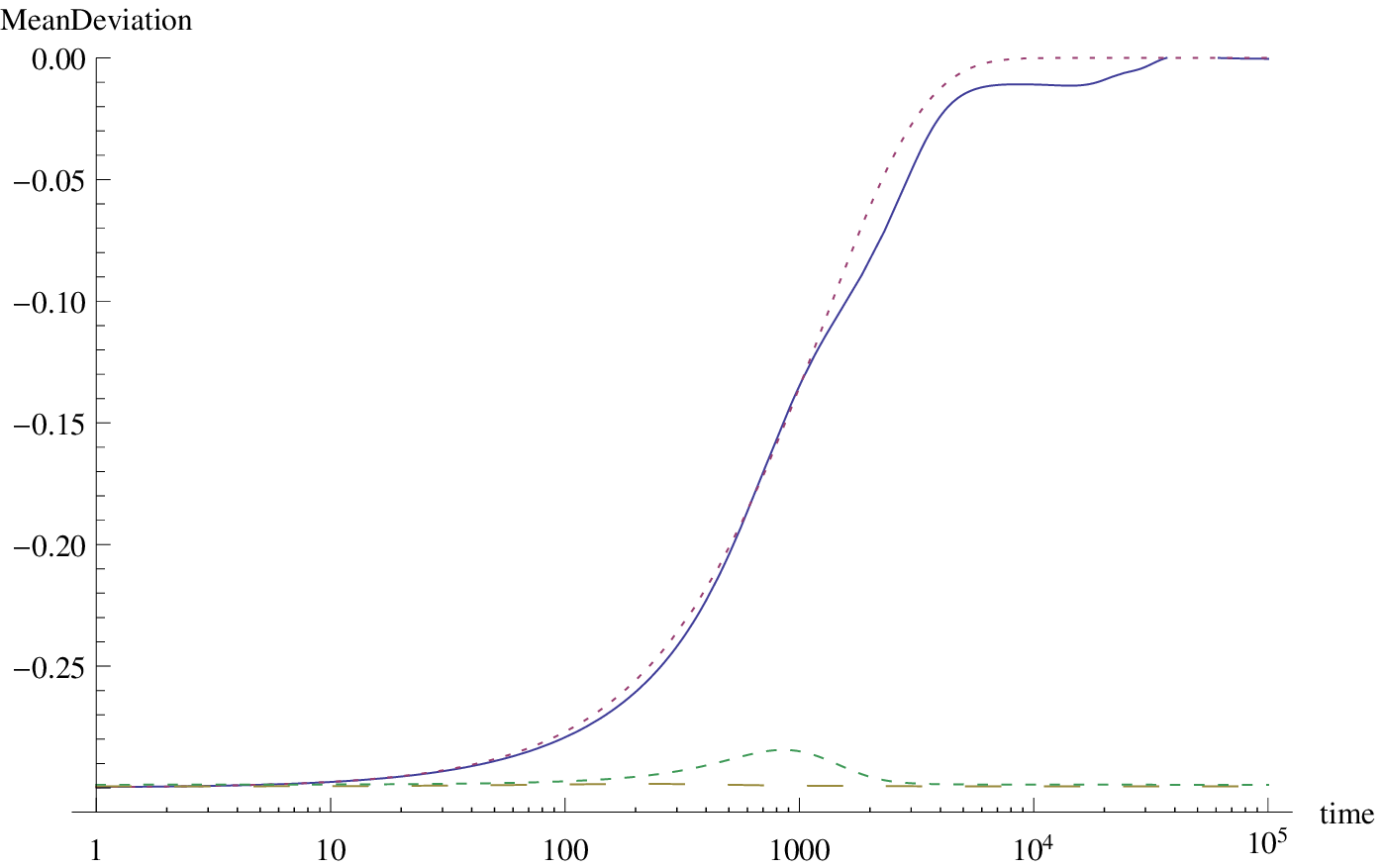}
\includegraphics[width=0.8\textwidth,angle=0]{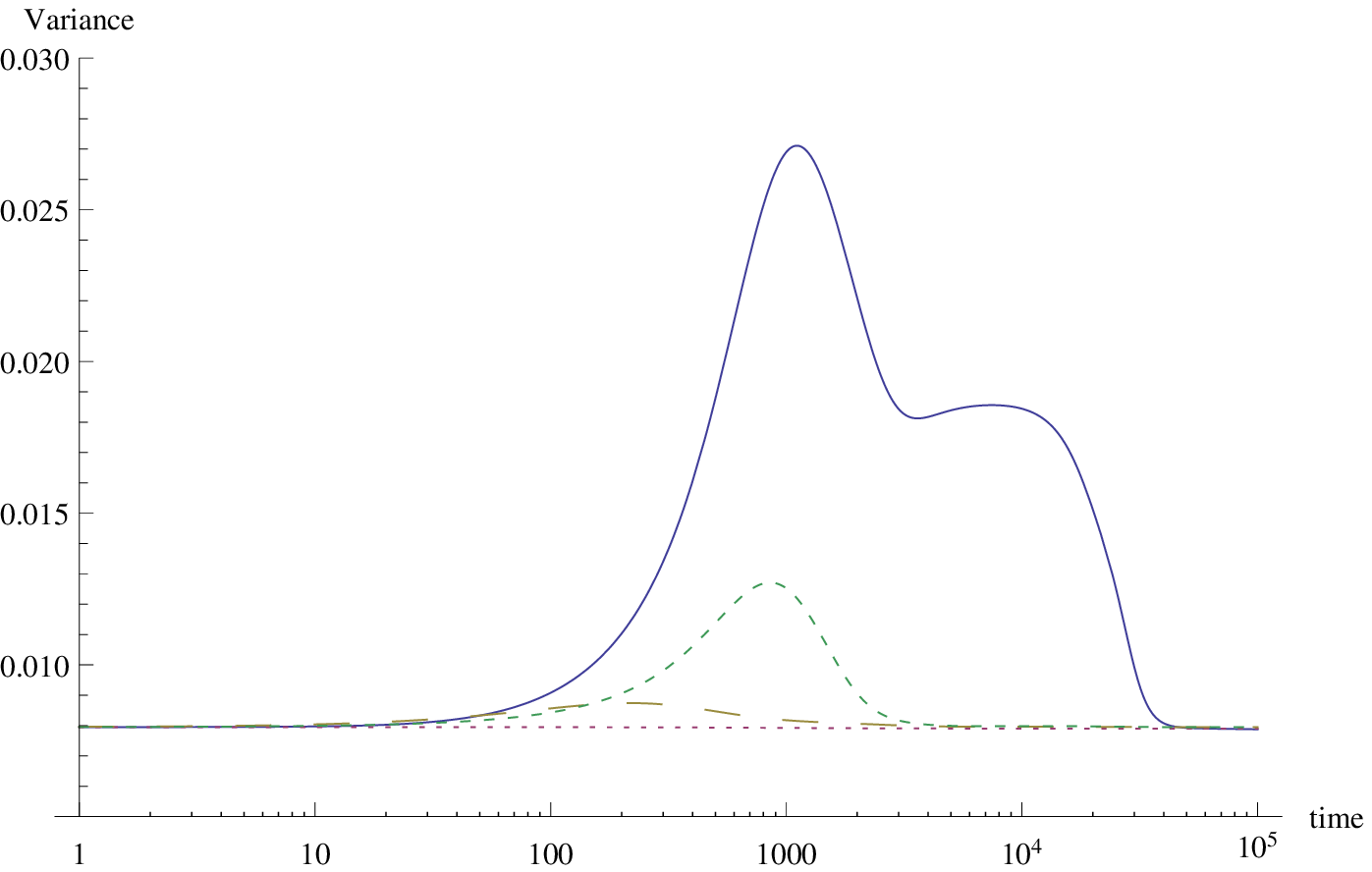}
 \end{center}
\caption{Response to change in optimum when most effects are large. 
  Solid lines show the mean deviation (\ref{meandefn}) and variance
  (\ref{vardefn}), while the two dashed curves show  the
  contribution to these cumulants from the first two relevant loci
  with effect $0.77$ (large dashes) and $0.34$ (small dashes). In
  both cases, the exact 
numerical solution of the full model is used. The dotted curves show 
(\ref{CHsoln2}) and (\ref{valid}) with ${\cal C}=0.006$. The final optimum
value $z_f=0.3$ and the other parameter values are the same as in
Fig.~\ref{largeres1}.} 
\label{largeresS}
 \end{figure}


\end{document}